\documentclass[11pt]{article}
\usepackage{fa2023}
\usepackage{amsmath}
\usepackage{cite}
\usepackage{url}
\usepackage{graphicx}
\usepackage{float}
\usepackage{color}
\usepackage[utf8]{inputenc}
\graphicspath{{C:/Users/user/Documents/MATLAB/PhD_programs/Application_biblio/ode45 complex modes NL/pdf_ForumAcusticum}}

\title{Modal decomposition synthesis for localized nonlinear losses at termination of a woodwind model: influence on sound characteristics}

\multauthor
{N. Szwarcberg$^{1^*,2}$ \hspace{1cm} T. Colinot$^1$ \hspace{1cm} C. Vergez$^2$} { \bfseries{M. Jousserand$^1$}\\
  $^1$ Buffet Crampon, 5 Rue Maurice Berteaux, 78711 Mantes-la-Ville, France\\
$^2$ Aix Marseille Univ, CNRS, Centrale Marseille, LMA, Marseille, France\\
\correspondingauthor{szwarcberg@lma.cnrs-mrs.fr}{Szwarcberg et al.}
}

\begin{document}
\maketitle
\begin{abstract}
A sound synthesis model for woodwind instruments is developed using modal decomposition of the input impedance, accounting for viscothermal losses as well as localized nonlinear losses at the end of the resonator. 
To extend the definition of the input impedance to the nonlinear domain, the method incorporates a dependence on the RMS acoustic velocity at a geometric discontinuity. 
The poles and residues resulting from the modal decomposition are then fitted with respect to this velocity.
Thus, the pressure-flow relation defined by the resonator is completed by new equations which account for the dependence with the RMS velocity at the end of the tube.
The ability of the model to reproduce the bifurcation diagram of a reed instrument was confirmed in a previous article from the authors \cite{szwarcberg2023amplitude}.
The present work focuses on the influence of localized nonlinear losses on the acoustic pressure signal.
Results are discussed with the observations of \cite{guillemain2006digital}.
\end{abstract}

\section{Introduction}\label{sec:introduction}

Sound synthesis by modal decomposition is a method which relies on the measurement or computation of the input impedance of the resonator.
This method allows to catch the geometric subtleties that differentiate two wind instruments.
However, the input impedance is a measure of the passive response of the resonator.
It does not account for phenomena related to high acoustic levels, which can occur in regular wind instrument playing conditions \cite{bergeot2014response}.

This article focuses on localized nonlinear losses.
They appear at the level of a geometrical discontinuity, such as a side hole \cite{ingaard1950acoustic} or the open end of the resonator \cite{disselhorst1980flow}.
In waveguide synthesis, they are taken into account by adding a boundary condition derived from Bernoulli's law at the discontinuity \cite{dalmont_oscillation_2007,atig2004saturation,guillemain2006digital,taillard2018phd}.

Moreover, Atig \cite{atig2004saturation, atig2004termination} and Dalmont \cite{dalmont_experimental_2002, dalmont_oscillation_2007} measured localized nonlinear losses in the form of a real impedance, depending on the amplitude of the acoustic velocity at the discontinuity as well as a loss coefficient, related to the geometry.
Diab et al. \cite{diab2022nonlinear} proposed a method to implement these nonlinear losses in an augmented input impedance.

Recently, the authors applied this method to a simplified clarinet consisting of a cylindrical resonator with nonlinear losses at the open end \cite{szwarcberg2023amplitude}. 
The proposed model manages to reproduce the experimental results of Atig \cite{atig2004saturation} and Dalmont \cite{dalmont_oscillation_2007}, concerning the saturation and extinction thresholds of a simplified clarinet.
This paper is a continuation of the latter.

Although nonlinear losses play a major role in the dynamic behavior of a wind instrument \cite{dalmont_oscillation_2007}, their influence on the timbre remains not well known \cite{atig2004phd}.
Delay line simulations including nonlinear losses \cite{guillemain2006digital} have described differences in the balance of the amplitude between even and odd harmonics during a \textit{crescendo}.

The objective of this paper is to investigate the potential effects on the spectral content induced by localized nonlinear losses at the open end of a simplified clarinet.
After presenting the model, comparisons of the spectral content will be carried out.

\section{Model}
\subsection{Resonator}
\subsubsection{Input impedance}
A simplified clarinet is studied.
The resonator is a cylindrical tube of length $L$ and cross-section $S= \pi R^2$, where $R$ is the bore radius. 
Nonlinear losses are taken into account at the open end of the tube through an impedance $Z_t$, given by \cite{atig2004termination} as:
\begin{equation}
Z_t = \frac{4 c_d}{3 \pi} \frac{v_{\mathrm{RMS}}}{c_0} Z_c,
\end{equation}
where $c_d \in[0,3]$ is a loss coefficient depending on the geometry of the open end \cite{atig2004saturation}, $v_{\mathrm{RMS}}$ is the RMS acoustic velocity at the discontinuity \cite{diab2022nonlinear} and $c_0$ is the speed of sound in the air.
The characteristic impedance of plane waves in the resonator is $Z_c= \rho c_0 /S$, where $\rho$ is the density of air.

The resistive impedance $Z_t$ is located in series with the radiation impedance $Z_R$, given by \cite[Chap. 12.6.1.3]{bible2016} for an unflanged pipe as:
\begin{equation}
Z_R =Z_c \left( jk\Delta l + \frac{1}{4}(kR)^2 \right),
\end{equation}
where $k=\omega /c_0$ is the wave number and $\Delta l = 0.6 R$.
The input impedance $Z_{\mathrm{in}}$ is therefore written as:
\begin{equation}\label{eq:Zin}
Z_{\mathrm{in}} =Z_c \tanh \left[ \Gamma L + \tanh^{-1} \left(\frac{Z_R + Z_t}{Z_c} \right) \right],
\end{equation}
where $\Gamma = jk + (1+j)\eta \sqrt{f}/R$ is the complex wave number for simplified viscothermal losses, $\eta=3 \cdot 10^{-5}$~s$^{1/2}$, and $f=\omega/(2\pi)$ is the frequency. 

\begin{figure}[h!]
	\centering	
	\includegraphics[width = .45\textwidth]{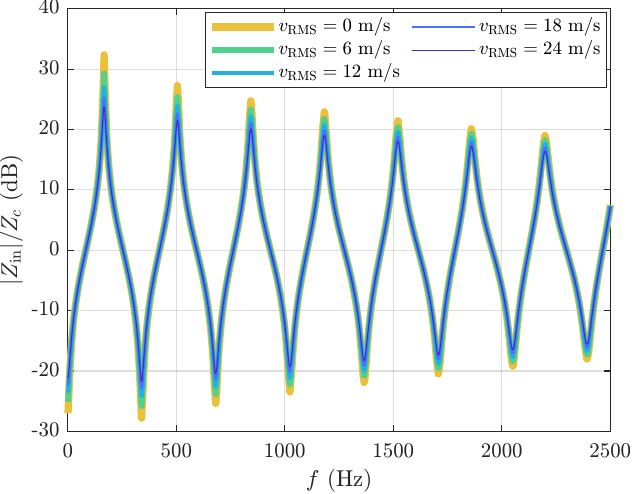}
	\caption{Input impedance $Z_{\mathrm{in}}$ of a cylinder of dimensions $L=50$~cm, $R=8$~mm, and $c_d=2.8$, corresponding to the experimental data of \cite{dalmont_oscillation_2007}. $Z_{\mathrm{in}}$ is plotted for increasing values of the RMS acoustic velocity at the discontinuity.}
	\label{fig:1}
\end{figure}

The evolution of the input impedance with respect to $v_{\mathrm{RMS}}$ is represented on Figure~\ref{fig:1}.
Nonlinear losses localized at the open termination have an influence on the amplitude of the resonance peaks, but not on their frequency.
This is accentuated for low frequencies. 
For $v_{\mathrm{RMS}}=24$~m$/$s, the amplitude of the first peak is $66\%$ lower than in the case without nonlinear losses, whereas for the fifth peak its decrease is reduced to $32\%$.

\subsubsection{Modal decomposition of the input impedance}

To be integrated in a physical model, the input impedance defined in Eq. \eqref{eq:Zin} must be decomposed into a sum of $N$ complex modes.
Each $n$-th mode is characterized by two complex coefficients: the poles $s_n$ and residues $C_n$.

To include nonlinear losses, \cite{diab2022nonlinear} express $C_n$ and $s_n$ with respect to $v_{\mathrm{RMS}}$.
The modal decomposition of $Z_{\mathrm{in}}$ is therefore written as:
\begin{multline} \label{eq:Zinmod}
Z_{\mathrm{in}}^{\mathrm{(modal)}} = Z_c \sum_{n=1}^N \frac{C_n(v_{\mathrm{RMS}})}{s - s_n(v_{\mathrm{RMS}})} \\
   + \frac{C_n^*(v_{\mathrm{RMS}})}{s - s_n^*(v_{\mathrm{RMS}})},
\end{multline}
where $s$ is the Laplace variable and $C_n^* = \mathrm{conj}(C_n)$.
The numerical computation of the modal coefficients with respect to $v_{\mathrm{RMS}}$ is detailed in \cite{szwarcberg2023amplitude}.
The evolution of the poles and residues is then fitted in the complex plane with respect to $v_{\mathrm{RMS}}$. 
It appears that linear regression is sufficient to describe their evolution:
\begin{align}
C_n(v_{\mathrm{RMS}}) &\approx C_n^{(0)} +  C_n^{(1)} v_{\mathrm{RMS}}, \\
s_n(v_{\mathrm{RMS}}) &\approx s_n^{(0)} + s_n^{(1)} v_{\mathrm{RMS}},
\end{align}
where the superscript $(0)$ denotes the value of the modal coefficient when $v_{\mathrm{RMS}}=0$, i.e. without nonlinear losses.
As the loss coefficient $c_d$ increases, the slopes $C_n^{(1)}$ and  $s_n^{(1)}$ steepen almost linearly.

\subsubsection{Equation for the physical model}
The modal decomposition formalism given by Eq. \eqref{eq:Zinmod} allows to write the following time-domain relation between the dimensionless input pressure $p(t)$ and flow $u(t)$:
\begin{align}\label{eq_zin_t}
\dot{p}_n(t) - s_n(v_{\mathrm{RMS}}) p_n(t) = C_n(v_{\mathrm{RMS}}) u(t), \\
\mathrm{where} \quad  p(t) = 2 \Re \left( \sum_n^N p_n(t) \right).
\end{align}
At each time step, the value of the poles and residues are updated after computation of $v_{\mathrm{RMS}}$ at the position where localized nonlinear losses occur. 
To do so, the pressure at the open end is derived from $p_n(t)$ through the mode shapes of the resonator. 
The time derivative of the acoustic velocity is then deduced from Euler's equation. 
The RMS acoustic velocity is finally computed by double time integration.
Further details are provided in \cite{szwarcberg2023amplitude}.

\subsection{Reed dynamics}
The reed is considered as a single degree-of-freedom oscillator, of angular resonance frequency $\omega_r=2\pi \cdot 2200$~rad$/$s and damping $q_r=0.4$. 
The equation for the dimensionless displacement $x(t)$ is given by:
\begin{equation}
\dfrac{1}{\omega_r^2}\ddot{x}(t) +\dfrac{q_r}{\omega_r}\dot{x}(t)+ x(t)=p(t)- \gamma(t),
\end{equation}
where $\gamma(t)$ is the dimensionless blowing pressure.
A "ghost reed" model is considered here \cite{colinot2019ghost}, i.e. the reed can go through the mouthpiece.

\subsection{Reed channel}
The flow $u(t)$ through the reed channel is a dimensionless version of the one given by \cite{wilson_operating_1974} as:
\begin{multline}\label{eq:flow}
u(t)=-\lambda \dot{x}(t) +  \zeta \left[x(t)+1\right]^+ \cdot \\
\mathrm{sgn} \bigl[\gamma(t)-p(t)\bigr] \sqrt{\left|\gamma(t)-p(t)\right|},
\end{multline}
where the superscript $^+$ denotes the positive part.
$\zeta=0.27$ is the embouchure parameter, and $\lambda=5.5\cdot 10^{-3}/c_0$ is the reed flow parameter \cite{chabassier2022control,dalmont2003nonlinear}.

\section{Simulation and results}
Simulations are performed from the experimental data of \cite{dalmont_oscillation_2007}.
The resonator used during the experiments was a cylindrical tube of length $L=50$~cm and radius $R=8$~mm.
The loss coefficient at the open end was estimated to $c_d=2.8$. 
The authors measured the dynamic thresholds of the clarinet for several embouchure configurations. 
In the present paper, an intermediate value of the embouchure parameter was arbitrarily chosen, such that $\zeta=0.27$.
Simulations were performed with eight modes, using the numerical solver \texttt{ode45} from \textsc{Matlab}.

In the following results, the dimensionless input pressure $p(t)$ was recorded for two different configurations.
The first one is a \textit{crescendo}, represented by a linear increase from $\gamma=0$ to $\gamma=3$ in 8~s.
The second one focuses on steady-state oscillations in standard playing conditions \cite{guillemain2006digital}, with $\gamma=0.56$.

The influence on nonlinear losses on the spectral content of the input pressure is investigated by comparing two resonators: including nonlinear losses at the open end ($c_d=2.8$) and not ($c_d=0$).

\subsection{Increasing blowing pressure}
\begin{figure}[h!]
	\centering	
	\includegraphics[width=.45\textwidth]{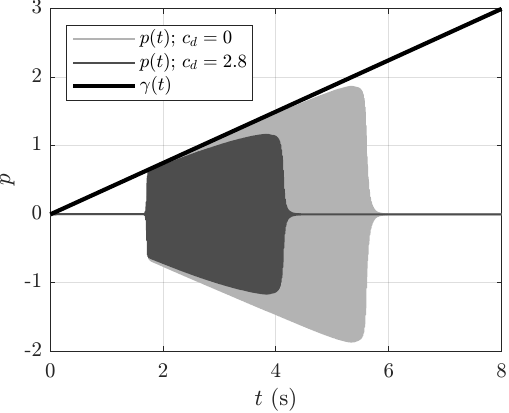}
	\caption{Evolution of the input pressure $p(t)$ for a linear increase of $\gamma(t)$. Light and dark-colored curves represent cases without and with nonlinear losses respectively.}
	\label{fig:2}
\end{figure}

Figure \ref{fig:2} shows the evolution of the pressure signal for a \textit{crescendo} of blowing pressure.
The impact of nonlinear losses on the extinction threshold (i.e. maximum blowing pressure for which  sound is still produced) is documented \cite{atig2004saturation,dalmont_oscillation_2007}.
Here, the focus is on the spectral content around the saturation of the resonator with nonlinear losses, which is around $t=3.8$~s.

\begin{figure}[h!]
	\centering	
	\includegraphics[width=.45\textwidth]{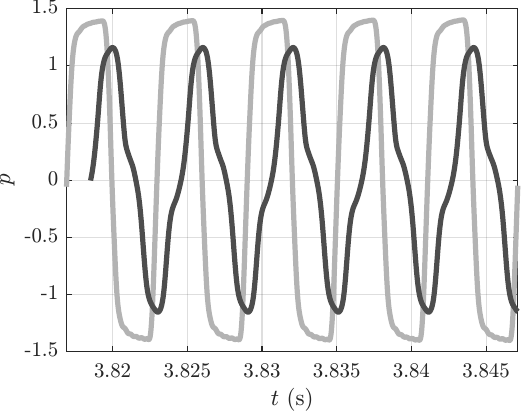}
	\includegraphics[width=.46\textwidth]{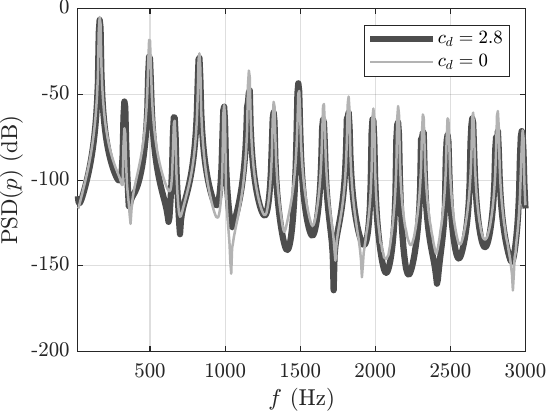}
	\caption{Detailed view on the pressure signal at the saturation of the resonator with nonlinear losses, around $\gamma\approx 1.23$.
	 The color code is the same as in Figure \ref{fig:2}. Top panel: time signal. Bottom panel: Power Spectral Density. Signals were normalized before processing for easier comparisons in the frequency domain.}
	\label{fig:3}
\end{figure}

Figure \ref{fig:3} represents the pressure signal around saturation in the time and frequency domains. 
For $c_d=0$ (no losses, light-colored curve), the pressure signal is close to a square wave, suggesting that odd harmonics are predominant over even harmonics.
In comparison, for $c_d=2.8$, the waveform suggests that even harmonics have a higher amplitude than in the previous case.

This is confirmed on the PSD spectrum represented on the bottom panel of Figure \ref{fig:3}.
From 1600 Hz, in the lossless case, the amplitudes of the harmonics are all higher than in the case with nonlinear losses.
Overall, the odd harmonics are more present without nonlinear losses than with, except for the fifth harmonic of the tube (the fifth odd harmonic of the signal) whose amplitude is slightly higher in the case with losses.
However, the first three even harmonics have higher amplitude with nonlinear losses than without.
Finally, the minima between peaks are overall of lower amplitude when nonlinear losses are taken into account.

The previous observations are in agreement with \cite{guillemain2006digital}, who point out that localized nonlinear losses at the open end have an influence on the balance between the amplitudes of even and odd harmonics.

This case corresponds to the configuration for which nonlinear losses have the most influence. 
Around the saturation pressure, $v_{\mathrm{RMS}}$ is maximal ($14.4$~m$/$s was calculated for $c_d=2.8$). 
The values of the modal coefficients are the most deviated from the case without losses, which can participate in the differences noted between the two spectra from Figure~\ref{fig:3}.
Moreover, the two systems are in very different situations, since the one including losses is close to its extinction threshold, while the other is in the middle of its dynamic range.
These dynamic differences may also have an influence on spectral discrepancies between the two signals.

\subsection{Steady state oscillations}
To minimize the differences potentially linked to the dynamic range of the systems, the pressure signals of the two resonators are compared for a constant and relatively small blowing pressure of $\gamma=0.56$.

\begin{figure}[H]
	\centering	
	\includegraphics[width=.45\textwidth]{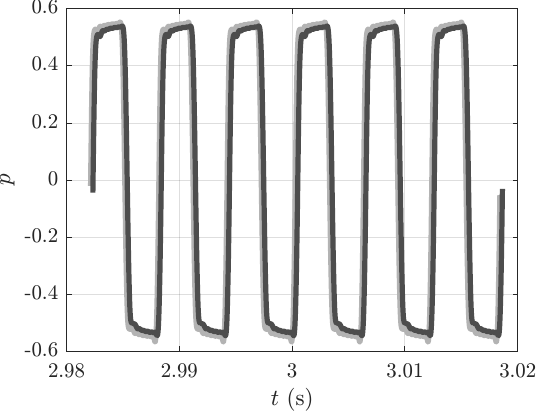}	
	\includegraphics[width=.45\textwidth]{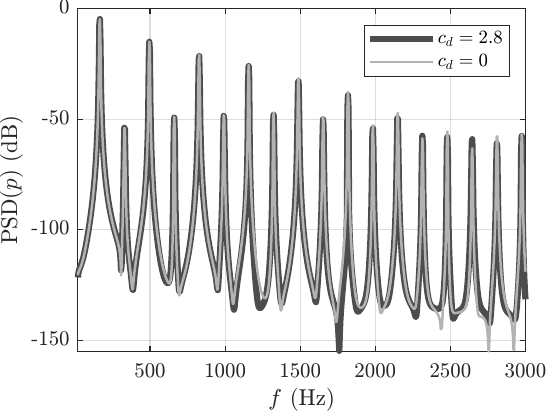}
	\caption{Pressure signal in the steady-state case, for $\gamma=0.56$. Top panel: time signal. Bottom panel: Power Spectral Density. Signals were normalized before processing for easier comparisons in the frequency domain.}
	\label{fig:4}
\end{figure}

Results are presented on Figure \ref{fig:4}.
The waveforms and the spectra are almost superimposed. 
The amplitude of $p(t)$ is $2.6\%$ higher when nonlinear losses are ignored.
Moreover, the spectra are almost identical up to about 1200~Hz.
From this frequency, the odd harmonics resulting from the lossless resonator are slightly higher.
Around 2500~Hz, the amplitude of the even harmonics are noticeably higher when losses are accounted for.

At a relatively low blowing pressure, the timbre differences induced by nonlinear losses located at the open end can reasonably be considered indistinguishable.
However, in these simulations, the value of $v_{\mathrm{RMS}}$ was calculated to $8.0$~m$/$s.
At this velocity, the input impedance defined by Eq. \eqref{eq:Zin} is distinctly different from the input impedance without losses (highest amplitude curve from Figure \ref{fig:1}).
Thus, this lack of spectral differences is rather unexpected.

\section{Conclusion}
Simulations were carried out to study the influence of nonlinear losses localized at the end of a tube on the input pressure produced by a clarinet-like system.
Attention was given to the spectral content in two different configurations.
First, comparisons were made near the saturation threshold of the system which included nonlinear losses. 
Overall, the amplitudes of the harmonics produced by the former system are smaller than in the case without nonlinear losses.
The second configurations focused on steady-state oscillations for a relatively small blowing pressure.
The two pressure profiles appear to be almost identical.
As a result, localized nonlinear losses at the open end are decisive to reproduce the dynamical behavior of a reed instrument, but based on the simulations carried out, their impact on sound characteristics can be considered as secondary. 
These conclusions should not be extrapolated to localized nonlinear losses in side holes, where the acoustic velocity can be higher due to the smaller cross-sectional area, and where some resonator modes may be more affected than others.
\section{Acknowledgments}
This study has been supported by the French ANR LabCom LIAMFI (ANR-16-LCV2-007-01). 
The authors thank S. Maugeais and J.-P. Dalmont for their valuable comments.

\bibliography{biblio}

%
%
%

\end{document}